\documentclass[12pt]{iopart}

\usepackage{graphicx,color}% Include figure files
\usepackage{bm}% bold math

\begin{document}

\title[Tight-binding description of Landau levels of graphite in
  tilted magnetic fields]
{Tight-binding description of Landau levels of graphite in
  tilted magnetic fields}
% Force line breaks with \\
\author{Nataliya A. Goncharuk and Ludv\'{i}k Smr\v{c}ka}
\address{Institute of Physics, Academy of Science of the Czech
Republic,\nolinebreak[5] v.v.i.,\\ 
Cukrovarnick\'{a} 10, 162 53 Prague 6, Czech Republic}

\eads{\mailto{gonchar@fzu.cz},\mailto{smrcka@fzu.cz}}


\begin{abstract}
The electronic structure of Bernal-stacked graphite subject to a tilted
magnetic field is studied theoretically. The minimal nearest-neighbor
tight-binding model with the Peierls substitution is employed to
describe the structure of Landau levels. We show that while the
orbital effect of the in-plane component of the magnetic field is
negligible for massive Dirac fermions in the vicinity of a the $K$ point
of the graphite Brillouin zone, at the $H$ point it leads to the experimentally
observable splitting of Landau levels, which grows approximately
linearly with the in-plane field intensity.
\end{abstract}

\pacs{71.20.-b, 71.70.Di}
\noindent{\it Keywords\/}: graphite electronic structure, tilted
magnetic field, splitting of Landau levels.

\maketitle

%\tableofcontents

%******* 
\section{Introduction}
\label{Introduction}
%*******
The recent attention paid to graphene monolayers has been motivated by
their unusual two\discretionary{-}{-}{-}dimensional (2D) Dirac energy
spectrum of electrons. In Bernal-stacked  graphene multilayer
composed of weakly coupled graphene sheets, the interlayer interaction
converts the 2D electron energy spectrum of graphene into the
three-dimensional (3D) spectrum of graphite. The electronic structure
of 3D graphite subject to magnetic fields perpendicular to $x-y$
planes of graphene layers were extensively studied a long time ago,
see, e.g., Refs.~\cite{McClure_1956,McClure_1960,Inoue_1962,Wallace_1972,
Dresselhaus_1974,Nakao_1976}.

The application of the tilted magnetic field $\vec{B}=(0, B_y, B_z)$
is a standard method used to distinguish between 2D and 3D electron
systems, as in 3D systems the orbital effect of the in-plane magnetic
field component should be observable. We will study this problem
theoretically using a simple tight-binding quantum mechanical model of
the graphite electron structure.

Various approaches were employed previously to study the influence of
the tilted magnetic fields.

The Fermi surfaces of metal single crystals were investigated by
measurements of the de Haas-van Alphen effect in tilted magnetic
fields. The interpretation of experiments relies on the quasiclassical
Onsager-Lifshitz quantization rule \cite{Onsager_1952,Lifshitz_1956},
the Fermi surface is reconstructed from the periods of
magneto-oscillations which are proportional to angular-dependent
extremal-cross-sections perpendicular to the direction of the tilted
magnetic field.

In semiconductor superlattices the quasiclassical interpretation of
data measured in tilted magnetic fields fails, as reported in
Refs.~\cite{Chang_1982,Stormer_1986,Jaschinski_1998,Nachtwei_1998,
Kawamura_2001,Goncharuk_2007}. In these papers the observed quantum
effects are attributed to the shift of centers of $k$-space orbits in
neighboring quantum wells by $|e|B_y/d$ in the $k_x$-direction, where
$d$ is the distance between quantum wells. In real space this means
that the in-plane magnetic-field length $\ell_y=\sqrt{\hbar/|e|B_y}$
should become comparable with $d$ to reach the visible
effect~\cite{Dingle_1978}.

Besides semiconductor superlattices, other layered materials with much
shorter interlayer distances were also investigated in tilted magnetic
fields. Different versions of angular magnetoresistance oscillations
(AMRO) were studied both experimentally and theoretically in
low-dimensional quasi-2D and quasi-one-dimensional organic conductors
(see, e.g.,Refs.~\cite{Kartsovnik_1988,Kajita_1989,Yamaji_1989,Yagi_1990,
Lebed_1989,Osada_1991,Danner_1994,Chashechkina_1998,Lee_1998} and
references therein), and also in intercalated graphite
\cite{Iye_1994,Enomoto_2006}. On the theory side, the high Landau
level (LL) filing factors and weak interlayer interaction were
considered in iterpretation the data.

In pristine graphite this problem has been touched on by two recent
theoretical articles.

The graphene multilayer energy spectrum in magnetic fields parallel to
the layers was described quantum-mechanically in
Ref.~\cite{Pershoguba_2010}, as the standard theory of AMRO in tilted
magnetic fields was not applicable due to the relatively strong
interlayer interaction (in comparison with the  intercalated graphite)
between graphene sheets.

The LLs in the bilayer graphene in magnetic fields of arbitrary
orientations were calculated analytically in Ref.~\cite{Hyun_2010}.

Both papers conclude that a very strong in-plane field component is
necessary to induce an observable effect on the electronic structure.
Indeed, to reach $\ell_y$ comparable with the distance between
graphene layers in graphite, the magnetic field $B_y = 5865$~T would
be necessary.

In this paper we make use of the specific features of the LL structure
in two nonequivalent neighboring graphene sheets in graphite, and show
that at the $H$ point of the graphite hexagonal Brillouin zone the
application of the tilted magnetic field leads to  experimentally
observable splitting of LLs of the order of several meV.

%*******
\section{Model}
\label{Model}
\begin{figure}[htb]
\centering
\includegraphics[width=0.85\linewidth]{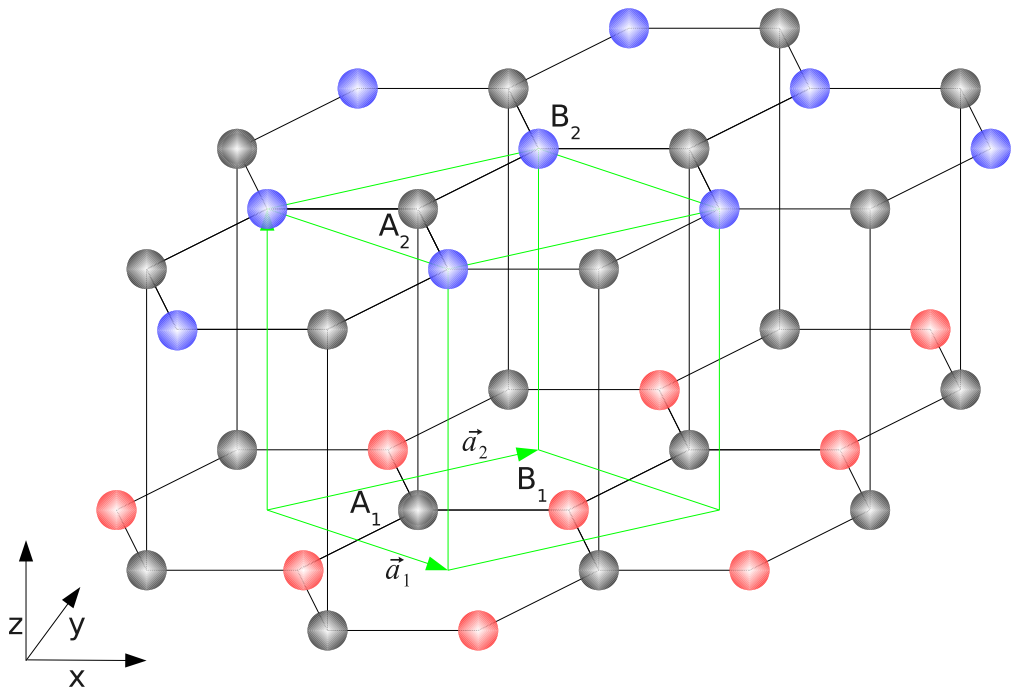}
\caption{(Color online) The lattice structure of graphite. 
The unit cell is a green parallelepiped.}
\label{Fig1}
\end{figure}

%*******
Bulk graphite is composed of periodically repeated graphene bilayers
formed by two nonequivalent Bernal-stacked graphene sheets, as shown
in Fig.~\ref{Fig1}. There are two sublattices, $A$ and $B$, on
each sheet and, therefore, four atoms in a unit cell. The distance
between the nearest atoms $A$ and $B$ in a single layer is $1.42\,$
{\AA}, the interlayer distance between nearest atoms $A$ is
$d=3.35\,${\AA}.

%//////////////////
%\\\\\\\\\\\\\\\\\\
 
To describe the graphite band structure, we employ the minimal
nearest-neighbor tight-binding model, introduced by Koshino
et al. in Ref.~\cite{Koshino_2008}. This model is reduced
Slonczewski, Weiss and McClure (SWM) model. 
The tight-binding Hamiltonian $\mathcal{H}$ includes only two instead
of seven tight-binding parameters, the intralayer
interaction $\gamma_0=3.16$~eV between the nearest atoms $A$ and $B$
in the plane, and the interlayer interaction $t=0.39$~eV between the
nearest atoms $A$ out of plane.

While the reduced SWM model is not appropriate, e.g., for a Fermi
surface description, this model has been successfully applied 
in the theoretical papers~\cite{Pershoguba_2010,Hyun_2010}, and 
used to describe recent magneto-optical measurements on graphite 
in Refs.~\cite{Henriksen_2008,Nicholas_2009,Orlita_2009,
Orlita_2010,Ubrig_2011}. It has been shown in 
Refs.~\cite{Orlita_2009,Orlita_2010,Ubrig_2011} 
that the transitions between Landau levels
originating from the $K$ and $H$ points of the graphite Brillouine zone can be
understood within a simple picture of an effective bilayer with a coupling
strength enhanced twice in comparison to a true graphene bilayer, $2t$,
(which is definitely the effect of a superlattice) and an effective graphene
monolayer. These solid arguments are in agreement with the theoretical model
based on the reduced SWM model we develop in our manuscript. 

In this model the
wave functions are expressed via four orthogonal components
$\psi^A_j$, $\psi^B_j$, $\psi^A_{j+1}$, $\psi^B_{j+1}$, which are, in
zero magnetic field, Bloch sums of atomic wave functions over the
lattice sites of sublattices $A$ and $B$ in individual layers $j$. 

The continuum approximation is used
in the vicinity of the $H - K - H$ axis of the graphite hexagonal
Brillouin zone, for small $\vec{k}=\left(k_x,k_y\right)$ measured from
the axis. Then the electron wave length is larger than the distance
between atoms, and the non-zero matrix elements of $\mathcal{H}$ can
be written as
\begin{eqnarray}
\mathcal{H}^{AB}&=\hbar v_F(k_x+ik_y),\\
\mathcal{H}^{BA}&=\hbar v_F(k_x-ik_y),\\
\mathcal{H}^{AA}&=t.
\end{eqnarray}
The Fermi velocity, $v_F$, is defined by $\hbar v_F
= \sqrt{3}a\gamma_0/2$, and will be used as an intralayer parameter
instead of $\gamma_0$ in the subsequent consideration.

The effect of the arbitrary oriented magnetic field, $\vec{B}=(0, B_y,
B_z)$, can be conveniently introduced into the zero-field Hamiltonian
by the Peierls substitution. If we choose the vector potential in the
Landau form $\vec{A}=(B_yz-B_zy, 0, 0)$, the substitution will read
\begin{equation}
\hbar k_x \rightarrow \hbar k_x-|e|B_zy+|e|B_yjd,
\label{eqkx}
\end{equation}
where an integer number $j$ indicates the graphite layer number.
Consequently, the
matrix elements, $\mathcal{H}^{AB}$ and $\mathcal{H}^{BA}$, 
become layer dependent in a tilted magnetic field,
\begin{eqnarray}
\mathcal{H}_j^{AB}=v_F(k_x-|e|B_zy+|e|B_yjd+ik_y)=v_F\Pi_j,
\label{eqhab1}\\
\mathcal{H}_j^{BA}=v_F(k_x-|e|B_zy+|e|B_yjd-ik_y)=v_F\Pi_j^{\ast}.
\label{eqhab2}
\end{eqnarray}

Making use of the above approximations, the Schr\"{o}dinger equation
involving all layers $j$, leads to the following system of equations
\begin{eqnarray}
v_F\Pi_{j}^{\ast}\psi_{j}^B-E\psi_{j}^A+
t\,\psi_{j-1}^A+t\,\psi_{j+1}^A=0,
\label{e1}\\
-E\psi_j^B+v_F\Pi_{j}^{\ast}\psi_{j}^A=0,
\label{e2}\\
v_F\Pi_{j+1}^{\ast}\psi_{j+1}^B-E\psi_{j+1}^A+
t\,\psi_{j}^A+t\,\psi_{j+2}^A=0,
\label{e3}\\
-E\psi_{j+1}^B+v_F\Pi_{j+1}\psi_{j+1}^A=0.
\label{e4}
\end{eqnarray}
Note that the structure of  $\mathcal{H}$ allows to express
the function $\psi_j^B$ via the function $\psi_j^A$ from the same
layer, and we are thus left with  two interlayer equations for $\psi_j^A$
and $\psi_{j+1}^A$
\begin{eqnarray}
(v_F^2\Pi_j \Pi_j^{\ast}-E^2)\psi_j^A+
t\, E(\psi_{j-1}^A+\psi_{j+1}^A )=0, 
\label{EQU1} 
\\
(v_F^2\Pi_{j+1}^{\ast} \Pi_{j+1}-E^2)\psi_{j+1}^A+
t\, E(\psi_{j}^A+\psi_{j+2}^A )=0.
\label{EQU2}
\end{eqnarray}
%

%*******
\section{Zero-field case}
%*******
It follows from the condition of periodicity in the $z$-direction that
$\psi_{j+1}^A$ and $\psi_j^A$ can be written as
\begin{equation}
\psi_{j+1}^A=e^{ik_zd(j+1)}\phi_1^A,\quad \psi_j^A=e^{ik_zdj}\phi_2^A,
\label{eqperio}
\end{equation}
where $\phi_1^A$ and $\phi_2^A$ denote the 2D wave functions in two
non-equivalent layers of the graphite unite cell, and $k_z$ is restricted to
the first Brillouin zone, $-\pi/2<k_zd<\pi/2$.

For $\vec{B}=0 $ Eqs.~(\ref{EQU1}, \ref{EQU2}) 
are transformed to
\begin{eqnarray}
(\hbar^2 v_F^2k^2 - E^2)\phi_2^A + 2 t\,E\cos(k_zd)\phi_1^A=0,\\
(\hbar^2 v_F^2k^2 - E^2)\phi_1^A + 2 t\,E\cos(k_zd)\phi_2^A=0,
\end{eqnarray}
and from here the four eigenvalues are obtained
\begin{equation}
E^{\pm,\pm} = \pm \mathcal{T} \pm \sqrt{\mathcal{T}^2+\hbar^2 v_F^2k^2},
\end{equation}
where 
\begin{equation}
\mathcal{T}=t\cos(k_zd)
\label{T} 
\end{equation}
denotes the $k_z$-dependent coupling of two graphene sheets.

%*******
\section{Perpendicular magnetic field}
%*******
In the perpendicular magnetic field, $B_y = 0$, the system remains
periodic in the $z$-direction, and Eq.~(\ref{eqperio}) is still
valid. To find a proper form of $\mathcal{H}^{AB}$ and
$\mathcal{H}^{BA}$ we introduce the perpendicular magnetic field
length, $\ell_z^2 = \hbar/(|e|B_z)$, the centre of a cyclotron orbit,
$y_0=\ell_z^2k_x$, the dimensionless variable, $\eta =
(y-y_0)/\ell_z$, and the perpendicular-magnetic-field-dependent
parameter, $\mathcal{B} = 2\hbar |e|B_zv_F^2$.  Then, in our notation,
\begin{eqnarray}
\mathcal{H}^{AB}&=v_F\Pi=-\sqrt{\mathcal{B}}\,a^{\dag}=
-\sqrt{\frac{\mathcal{B}}{2}}\left(-\frac{\partial}{\partial\eta}+
\eta\right),\\ 
\mathcal{H}^{BA}&=v_F\Pi^{\ast}=-\sqrt{\mathcal{B}}\,a=
-\sqrt{\frac{\mathcal{B}}{2}}
\left(\frac{\partial}{\partial\eta}+\eta\right),
\end{eqnarray}
$a^{\dag}$ and $a$ being raising and lowering operators, respectively.

\begin{figure}[htb]
\centering
\includegraphics[width=0.95\linewidth]{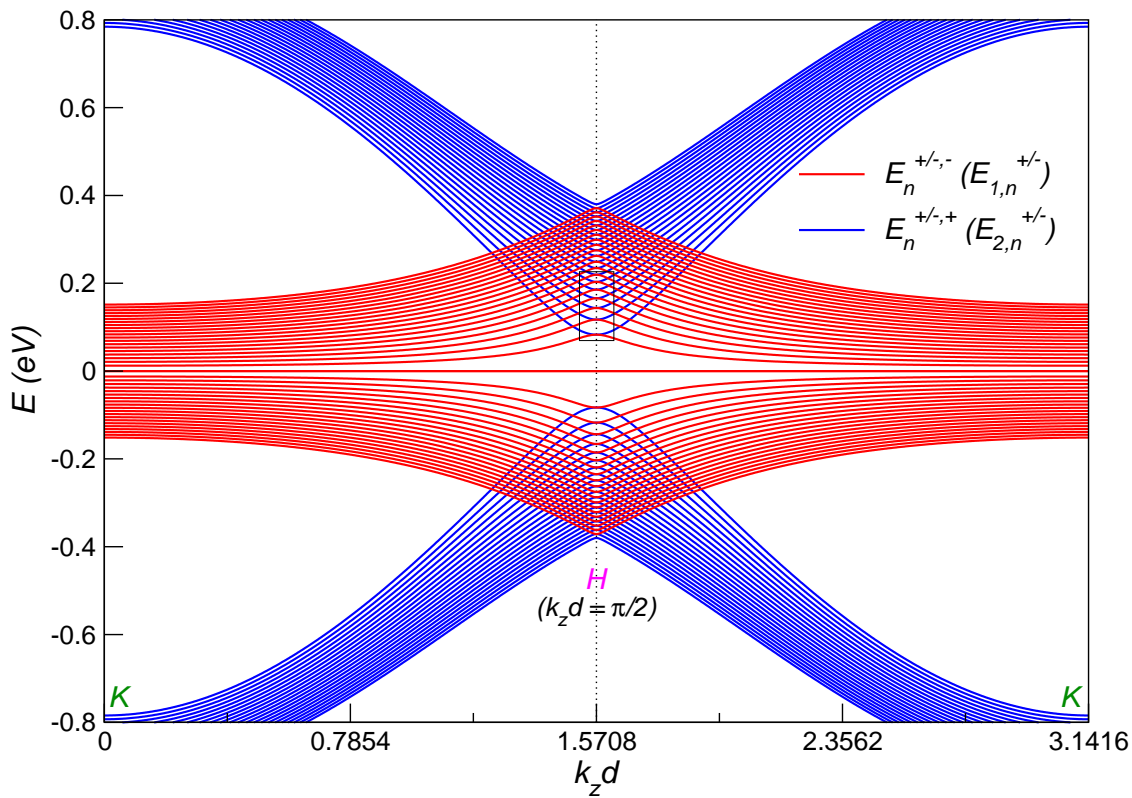}
\caption{(Color online) Landau subbands of graphite, $E_{n}^{\pm,-}$
and  $E_{n}^{\pm,+}$ (denoted as $E_{1,n}^{\pm}$  and $E_{2,n}^{\pm}$ 
at the $H$ point) subject to perpendicular magnetic field $B_z=5$~T, 
as a function of $k_z$ along the K-H-K pass of the Brillouin zone. The
$k_z$ dependence of LLs in the vicinity of the $H$ point restricted by
the rectangular box will be compared with those subject to the tilted
magnetic field, as shown in Fig.~\ref{Fig6}.}
\label{Fig2}
\end{figure}

With help of these expressions, Eqs.~(\ref{EQU1}) and (\ref{EQU2})
can be written as
\begin{eqnarray}
\left[\frac{\mathcal{B}}{2}\left(-\frac{\partial^2}{\partial\eta^2}
+\eta^2+1\right)-E^2\right]\phi_2^A + 2\mathcal{T}E\phi_1^A=0,
\label{eq1B}\\
\left[\frac{\mathcal{B}}{2}\left(-\frac{\partial^2}{\partial\eta^2}
+\eta^2-1\right)-E^2\right]\phi_1^A + 2\mathcal{T}E\phi_2^A=0.
\label{eq2B}
\end{eqnarray}
It is obvious from these equations that $\phi_1^A$ and $\phi_2^A$ are
closely related to the eigenfunctions of the harmonic oscillator,
$\varphi_n(\eta)$. Assuming
%\begin{eqnarray}
%\phi_1^A=\frac{1}{L_x}e^{ik_xx}\sum_{n'=0}^{\infty}A_{1,n'}
%\varphi_{n'}(\kappa),
%\nonumber\\
%\phi_2^A=\frac{1}{L_x}e^{ik_xx}\sum_{n'=0}^{\infty}A_{2,n'}
%\varphi_{n'}(\kappa),
%\end{eqnarray}
\begin{equation}
\phi_1^A=\frac{1}{L_x}e^{ik_xx}\sum_{n'=0}^{\infty}A_{1,n'}
\varphi_{n'}(\kappa),\quad
\phi_2^A=\frac{1}{L_x}e^{ik_xx}\sum_{n'=0}^{\infty}A_{2,n'}
\varphi_{n'}(\kappa),
\end{equation}
and having in mind that
\begin{equation}
\left(-\frac{\partial^2}{\partial\eta^2}
+\eta^2\right)\varphi_n(\eta)=(2n+1)\varphi_n(\eta),
\end{equation}
we get
\begin{eqnarray}
\left[\mathcal{B}(n+1)-E^2\right] A_{2,n}+
2 E \sum_{n'=0}^{\infty}\mathcal{T}_{n,n'}A_{1,n'}=0,\label{eq1p}
\\
\left[\mathcal{B}n-E^2\right] A_{1,n}+
2 E \sum_{n'=0}^{\infty}\mathcal{T}_{n,n'}A_{2,n'}=0,\label{eq2p}
\end{eqnarray}
where $T_{n,n'}$ is defined by
\begin{equation}
\mathcal{T}_{n,n'}=
\mathcal{T}\int_{-\infty}^{+\infty}\varphi_{n}(\eta)\varphi_{n'}(\eta)d\eta
=\mathcal{T}\delta_{n,n'}.
\label{t}
\end{equation}
It follows from (\ref{t}) that only the LLs with the same quantum numbers $n$
(but with the different energies) are coupled and we arrive to
\begin{eqnarray}
\left[\mathcal{B} (n+1)-E^2\right]A_{2,n}+2\mathcal{T}_{n,n}EA_{1,n}=0, 
\label{e1-Bz}\\
\left(\mathcal{B} n-E^2\right)A_{1,n}+2\mathcal{T}_{n,n}EA_{2,n}=0. 
\label{e2-Bz}
\end{eqnarray}
Solving Eqs.~(\ref{e1-Bz}, \ref{e2-Bz}) yields the eigenenergies
%\begin{eqnarray}
%E_{n}^{\pm,\pm}=&\pm
%\left\{2\mathcal{T}_{n,n}^2+\mathcal{B}\left(n+\frac{1}{2}\right)
%\pm\right.\nonumber\\
%&\left.\pm \left[4\mathcal{T}_{n,n}^4
%+4\mathcal{T}_{n,n}^2\mathcal{B}(n+\frac{1}{2})+
%\frac{\mathcal{B}^2}{4}\right]^{\frac{1}{2}}\right\}^{\frac{1}{2}}. 
%\end{eqnarray}
\begin{equation}
E_{n}^{\pm,\pm}=\pm
\sqrt{2\mathcal{T}_{n,n}^2+\mathcal{B}\left(n+\frac{1}{2}\right)
\pm \sqrt{4\mathcal{T}_{n,n}^4
+4\mathcal{T}_{n,n}^2\mathcal{B}(n+\frac{1}{2})+
\frac{\mathcal{B}^2}{4}}}, 
\end{equation}
which are presented in Fig.~\ref{Fig2}.
%//////////////////
%\\\\\\\\\\\\\\\\\\

As the densities of states of the above Landau subbands have
singularities at the points $K$ and $H$, $k_zd=0$ and $k_zd=\pi/2$,
respectively, we concentrate on the field dependence of levels
corresponding to these points.

At the $K$ point the eigenenergies are given by
%\begin{eqnarray}
%E_{n}^{\pm,\pm}=&\pm\left\{2 t^2+\mathcal{B}
%\left(n+\frac{1}{2}\right)\right.\pm
%\nonumber\\
%&\pm\left.\left[4 t^4+4t^2\mathcal{B}(n+\frac{1}{2})+
%\frac{\mathcal{B}^2}{4}\right]^{\frac{1}{2}}\right\}^{\frac{1}{2}},
%%E_{1(2),n}^{-}&=&-E_{1(2),n}^{+},
%\end{eqnarray}
\begin{equation}
E_{n}^{\pm,\pm}=\pm\sqrt{2 t^2+\mathcal{B}
\left(n+\frac{1}{2}\right)\pm\sqrt{4 t^4+4t^2\mathcal{B}(n+\frac{1}{2})+
\frac{\mathcal{B}^2}{4}}},
\end{equation}
i.e., the subbands are equivalent to those of a graphene bilayer with
the coupling constant doubled: $2t$ instead of~$t$.
%//////////////////
\begin{figure}[htb]
\centering
\includegraphics[width=0.95\linewidth]{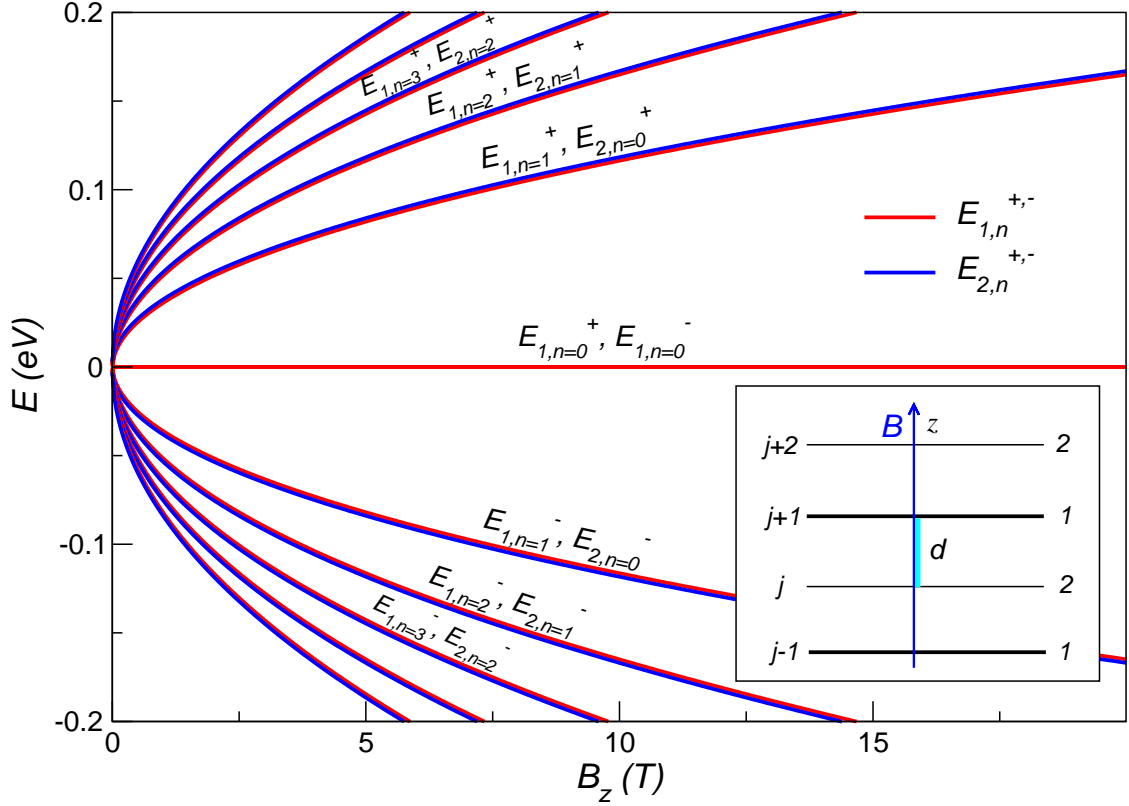}
\caption{(Color online) LLs of graphite at the $H$ point of the
Brillouin zone, $E_{1,n}^{\pm}$ and $E_{2,n}^{\pm}$, in the
perpendicular magnetic field, $B_z$.}
\label{Fig3}
\end{figure}
%\\\\\\\\\\\\\\\\\\

At the $H$ point the eigenenergies read
\begin{equation}
 E_{n}^{\pm,\pm}=\pm\sqrt{\mathcal{B}\left(n+\frac{1}{2}\right)\mp
  \frac{\mathcal{B}}{2}}.
\end{equation}
Due to the effectively vanishing inter-layer coupling the spectrum
corresponds to the Dirac fermions. The coefficients
$A_{1,n}$ and $A_{2,n}$ in Eqs.~(\ref{e1-Bz}, \ref{e2-Bz}) are equal
to $1$ and this implies that the corresponding wave functions are
localized either in the layer $1$ or in the layer $2$. To emphasize it, we
will write the eigeneneries as
\begin{eqnarray}
 E_{1,n}^{\pm}&=E_{n}^{\pm,-}=\pm\sqrt{\mathcal{B}n},\\
 E_{2,n}^{\pm}&=E_{n}^{\pm,+}=\pm\sqrt{\mathcal{B}(n+1)}.
\end{eqnarray}
The energy spectrum of LLs at the $H$ point is presented in Fig.~\ref{Fig3}. 

We also denote the wave functions $\phi_{1(2)}^A$ and $\phi_{1(2)}^B$
as $|n\rangle_{1(2)}^A$ and $|n\rangle_{1(2)}^B$ to stress that they
are the envelope wave functions of atomic wave functions $A$ and $B$
in the layers 1 and 2, as shown in TABLE~\ref{Table1}.

Let us mention that $E_{2,n}^{\pm}=E_{1,n+1}^{\pm}$, i.e., we are left
with pairs of degenerated LLs with different quantum numbers $n$ but
the same eigenenergies, with the wave functions localized in two
different layers.
%%%%%%%%%%%%%
\begin{table}[htb]
\caption{Energies and wave functions at the graphite $H$ point 
in perpendicular magnetic fields}
\begin{center}
\begin{tabular}{|c|c|}
\hline &\\ Energy & \{$\phi_1^A,\,\,\,\phi_1^B,\,\,\,\,
\phi_2^A,\,\,\,\,\phi_2^B$\}\\ \hline\hline
&\\ $E_1^+=\sqrt{\mathcal{B}n}$ & $\{|n\rangle_1^A,-|n-1\rangle_1^B,
0\,\,,0\,\}$\\ &\\ \hline 
&\\ $E_1^-=-\sqrt{\mathcal{B}n}$ &
$\{|n\rangle_1^A, \,\,\,|n-1\rangle_1^B,
0\,\,,0\,\}$\\ &\\ \hline 
&\\ $E_2^+=\sqrt{\mathcal{B}(n+1)}$ & 
$\{0,\,\,0\,,
|n\rangle_2^A,-|n+1\rangle_2^B\}$\\ &\\ \hline 
&\\ $E_2^-=-\sqrt{\mathcal{B}(n+1)}$
&$ \{0,\,\,0\,,
|n\rangle_2^A, \,\,\,|n+1\rangle_2^B\}$\\ &\\ \hline
\end{tabular}
\end{center}
\label{Table1}
\end{table}
%%%%%%%%%%%%%

%*******
\section{Tilted magnetic field}
%*******
While the previous two paragraphs summarized the already published
theories devoted to $\vec{B}=0$ and $\vec{B}= (0,0,B_z)$, here we present new
results for $\vec{B}= (0,B_y,B_z)$.

\begin{figure}[t]
\centering
\includegraphics[width=0.8\linewidth]{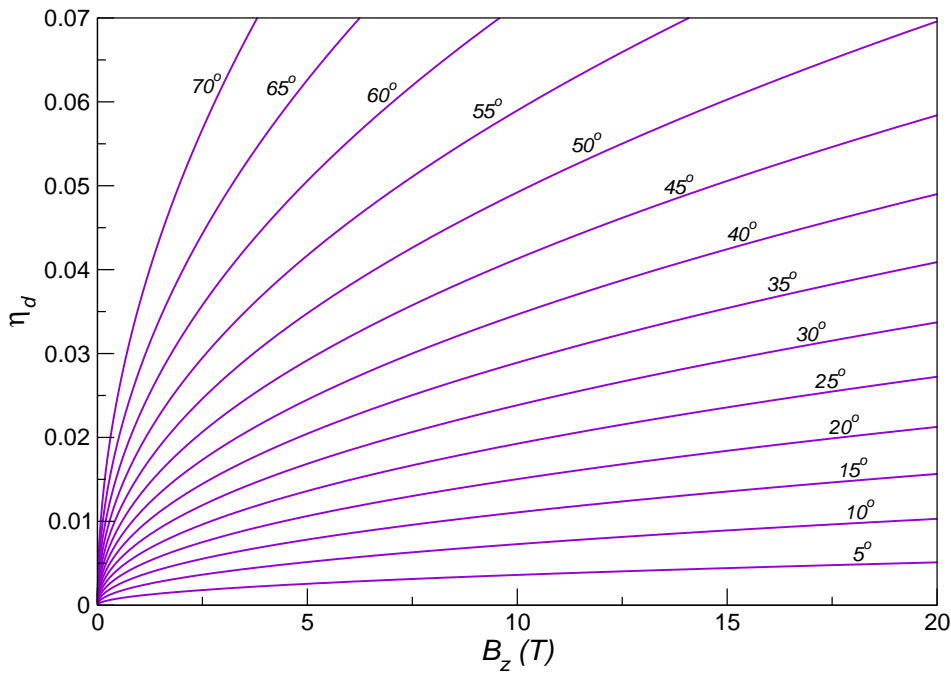}
\caption{(Color online) The dimensionless parameter $\eta_d$ as a
function of the perpendicular component of the magnetic field, $B_z$,
calculated for various tilt angles of the magnetic field $\vec{B}$. }
\label{Fig4}
\end{figure}

In tilted magnetic fields the off-diagonal matrix elements,
$\mathcal{H}_j^{AB}$ and $\mathcal{H}_j^{BA}$, given by
Eqs.~(\ref{eqhab1}, \ref{eqhab2}) remain layer dependent and take the form
\begin{eqnarray}
\mathcal{H}_j^{AB}&=v_F\Pi_j=-\sqrt{\frac{\mathcal{B}}{2}}
\left(-\frac{\partial}{\partial\eta}+\eta-j\eta_d\right),\\
\mathcal{H}_j^{BA}&=v_F\Pi_j^{\ast}=-\sqrt{\frac{\mathcal{B}}{2}}
\left(\frac{\partial}{\partial\eta}+\eta-j\eta_d\right),
\end{eqnarray}
where the small dimensionless parameter
\begin{equation}
\eta_d = \frac{B_y}{B_z}\frac{d}{\ell_z}
%= \frac{\ell_z d}{\ell_y^2}
\end{equation}
means the shift of the cyclotron orbit center in the $j$-layer due to
the in-plane component of the magnetic field,
$B_y$. Fig.~\ref{Fig4} illustrates the $B_z$ dependence of
$\eta_d$ for various tilt angles of the magnetic field.

Note that graphite subject to tilted magnetic fields is no longer
periodic in the $z$-direction, but becomes periodic in the direction
of the tilted magnetic field. To take into account the shift of the
cyclotron orbits, we apply the approach developed in
Ref.~\cite{Goncharuk_2005} for semiconductor superlattices.
%////////////////////////
%\\\\\\\\\\\\\\\\\\\\\\\\\\
Accordingly, the Eqs.~(\ref{EQU1}) and (\ref{EQU2}) are modified to
%\begin{eqnarray}
%\left[\frac{\mathcal{B}}{2}\left(-\frac{\partial^2}{\partial\eta^2}
%+\left(\eta-j\eta_d\right)^2+1\right)-E^2\right]\psi_j^A+\nonumber\\
%+t\, E\left(\psi_{j-1}^A+\psi_{j+1}^A\right)=0, 
%\label{eq1t}\\ 
%\left[\frac{\mathcal{B}}{2}\left(-\frac{\partial^2}{\partial\eta^2}
%+\left(\eta-(j-1)\eta_d\right)^2-1\right)-E^2\right]\psi_{j+1}^A+\nonumber\\
%+t\, E\left(\psi_{j}^A+\psi_{j+2}^A\right)=0.
%\label{eq2t}
%\end{eqnarray}
\begin{eqnarray}
\left[\frac{\mathcal{B}}{2}\left(-\frac{\partial^2}{\partial\eta^2}
+\left(\eta-j\eta_d\right)^2+1\right)-E^2\right]\psi_j^A+
t\, E\left(\psi_{j-1}^A+\psi_{j+1}^A\right)=0, 
\label{eq1t}\\ 
\left[\frac{\mathcal{B}}{2}\left(-\frac{\partial^2}{\partial\eta^2}
+\left(\eta-(j-1)\eta_d\right)^2-1\right)-E^2\right]\psi_{j+1}^A+\nonumber\\
+t\, E\left(\psi_{j}^A+\psi_{j+2}^A\right)=0.
\label{eq2t}
\end{eqnarray}

The new periodicity implies that $\psi_{j}^A$ and $\psi_{j+1}^A$ can be
written as
%\begin{eqnarray}
%\psi_{j}^A=&\, e^{i k_z d j}\phi_1^A\left(\eta+j\eta_d\right),\nonumber\\
%\psi_{j+1}^A=&\, e^{i k_z d(j+1)}\phi_2^A\left[\eta+(j+1)\eta_d\right],
%\label{eqperiot}
%\end{eqnarray}
\begin{equation}
\psi_{j}^A=\, e^{i k_z d j}\phi_1^A\left(\eta+j\eta_d\right),\quad
\psi_{j+1}^A=\, e^{i k_z d(j+1)}\phi_2^A\left[\eta+(j+1)\eta_d\right],
\label{eqperiot}
\end{equation}
where again $-\pi/2\leq k_z d\leq \pi/2$.  Here the wave functions
$\phi_1^A\left(\eta+j\eta_d\right)$ and
$\phi_2^A\left(\eta+j\eta_d\right)$ are associated with cyclotron
orbits in two layers.  Introducing the shift operator by
\begin{equation}
\phi\left(\eta+\eta_d\right)=e^{i\kappa\eta_d}\phi(\eta),\,\,\,\
\kappa=-i\partial/\partial\eta,
\end{equation}
and employing the $\kappa$-representation, two interlayer 
Eqs.~(\ref{eq1t}) and (\ref{eq2t}) can be given the form similar to 
Eqs.~(\ref{eq1B}, \ref{eq2B}) for the perpendicular magnetic field
\begin{eqnarray}
\left[\frac{\mathcal{B}}{2}\left(-\frac{\partial^2}{\partial\kappa^2}
+\kappa^2+1\right)-E^2\right]\phi_2^A+2\widetilde{\mathcal{T}}E\phi_1^A =  0 ,
\label{main-eq1}\\
\left[\frac{\mathcal{B}}{2}\left(-\frac{\partial^2}{\partial\kappa^2}
+\kappa^2-1\right)-E^2\right]\phi_1^A+2\widetilde{\mathcal{T}}E\phi_2^A =  0,
\label{main-eq2} 
\end{eqnarray}
but with the new coupling $\widetilde{\mathcal{T}}$ which depends, due to new
periodicity, not only on $k_z$ but also on both components $B_z$ and
$B_y$ of the arbitrary oriented magnetic field via $\eta_d$,
\begin{equation}
\widetilde{\mathcal{T}}(\kappa)=t\cos{(\kappa\eta_d+k_zd)}.
\label{T'}
\end{equation}

The Eqs.~(\ref{main-eq1}) and (\ref{main-eq2}) represent the main
result of this paper and in the following we will discuss the possible
methods of their solutions.

Expressing again $\phi_1^A$ and $\phi_2^A$  with a help of
(\ref{eq1p}) and (\ref{eq2p}) we arrive to
\begin{eqnarray}
\left[\mathcal{B}(n+1)-E^2\right] A_{2,n}+
2 E \sum_{n'=0}^{\infty}\widetilde{\mathcal{T}}_{n,n'}A_{1,n'}=0,\label{eq1pt}
\\
\left[\mathcal{B}n-E^2\right] A_{1,n}+
2 E \sum_{n'=0}^{\infty}\widetilde{\mathcal{T}}_{n,n'}A_{2,n'}=0,\label{eq2pt}
\end{eqnarray}
where
\begin{equation}
\widetilde{\mathcal{T}}_{n,n'}=
\int_{-\infty}^{\infty}\varphi_{n}(\kappa)\widetilde{\mathcal{T}}(\kappa)
\varphi_{n'}(\kappa)d\kappa.
\label{T'int}
\end{equation}
The integrals (\ref{T'int}) can be evaluated analytically and
expressed via the generalised Laguerre polynomials (see, e.g., Ref.~
\cite{Ryzhik}). The Eqs. (\ref{eq1pt}) and (\ref{eq2pt}) define the
matrix which should be diagonalized. The nonzero coupling of LLs with
different $n$ allows us to conclude that the degeneracy of LLs at the
$H$ point will be removed, and the LLs with different $n$ avoid to cross. 
Then the standard approach is to solve the secular
equation numerically, the minor complication being that the matrix
elements depend on the energy.

Here we prefer to obtain analytic results by the
lowest order perturbation theory application and treating $\eta_d$ as a small
parameter.

A simple trigonometric relation and a series expansion restricted to
terms linear in $\eta_d$ imply
\begin{eqnarray}
\widetilde{\mathcal{T}}(\kappa)&=t\cos(k_zd)\cos{(\kappa\eta_d)}-
t\sin{(k_zd)}\sin{(\kappa\eta_d)}\approx\nonumber\\
&\approx t\cos(k_zd) -t\sin{(k_zd)}\eta_d\kappa \cdots,
\label{T'-1}
\end{eqnarray}
and
%\begin{eqnarray}
%\widetilde{\mathcal{T}}_{n,n'} =& t\cos(k_zd)\delta_{n,n'}\nonumber\\
%&-t\eta_d\sin{(k_zd)}\int_{-\infty}^{\infty}\varphi_{n}(\kappa)
%\kappa\varphi_{n'}(\kappa)d\kappa.
%\label{T'int1}
%\end{eqnarray}
\begin{equation}
\widetilde{\mathcal{T}}_{n,n'} = t\cos(k_zd)\delta_{n,n'}-
t\eta_d\sin{(k_zd)}\int_{-\infty}^{\infty}\varphi_{n}(\kappa)
\kappa\varphi_{n'}(\kappa)d\kappa.
\label{T'int1}
\end{equation}
The  integrals in Eq.~(\ref{T'int1}) can be easily evaluated  using
the relation (see, e.g., Ref.~\cite{Blokhincev})
\begin{equation}
\kappa\varphi_{n}(\kappa)=\sqrt{\frac{n+1}{2}}\varphi_{n+1}(\kappa)+
\sqrt{\frac{n}{2}}\varphi_{n-1}(\kappa).
\end{equation}

Let us  pay attention to the field dependence of LLs  at
the most interesting $K$ and $H$ points of the graphite Brillouin zone.

At the $K$ point $k_zd=0$, and, consequently, $\cos(k_zd)=1$,
$\sin(k_zd)=0$.  Then $\widetilde{\mathcal{T}}_{n,n'}$ reduces to
\begin{equation} 
\widetilde{\mathcal{T}}_{n,n'} =t\delta_{n,n'}.
\end{equation}
This coupling corresponds to the effective bilayer subject to the perpendicular
field discussed above. Thus, we have found that corrections induced by
$B_y$ are very small and of the order $\eta_d^2$. This is in agreement
with conclusions presented  in  Ref.~\cite{Hyun_2010}.

The field dependence at the $H$ point, $k_zd=\pm \pi/2$, is more
interesting. In perpendicular magnetic fields the coupling between
layers disappears, and, as  presented above, we obtained the LLs 
corresponding to graphene
Dirac fermions, namely $E_{1,n}^{\pm}=\pm\sqrt{\mathcal{B}n}$ for the
first layer, and $E_{2,n}^{\pm}=\pm\sqrt{\mathcal{B}(n+1)}$ for the
second layer, $n=0,1,2,\cdots$.

In the magnetic field of an arbitrary direction the interlayer
interaction is not reduced to zero, but remains finite. The non-zero
matrix elements $\widetilde{\mathcal{T}}_{n,n'}$ can be written as
%\begin{eqnarray}
%\widetilde{\mathcal{T}}_{n,n+1}&=\widetilde{\mathcal{T}}_{n+1,n} = 
%t \eta_d \sqrt{\frac{n+1}{2}},\nonumber\\
%\widetilde{\mathcal{T}}_{n,n-1}&=\widetilde{\mathcal{T}}_{n-1,n} = 
%t \eta_d \sqrt{\frac{n}{2}}.
%\end{eqnarray}
\begin{equation}
\widetilde{\mathcal{T}}_{n,n+1}=\widetilde{\mathcal{T}}_{n+1,n} = 
t \eta_d \sqrt{\frac{n+1}{2}},\quad
\widetilde{\mathcal{T}}_{n,n-1}=\widetilde{\mathcal{T}}_{n-1,n} = 
t \eta_d \sqrt{\frac{n}{2}}.
\end{equation}

%////////////////////////
\begin{figure}[htb]
\centering
\includegraphics[width=0.95\linewidth]{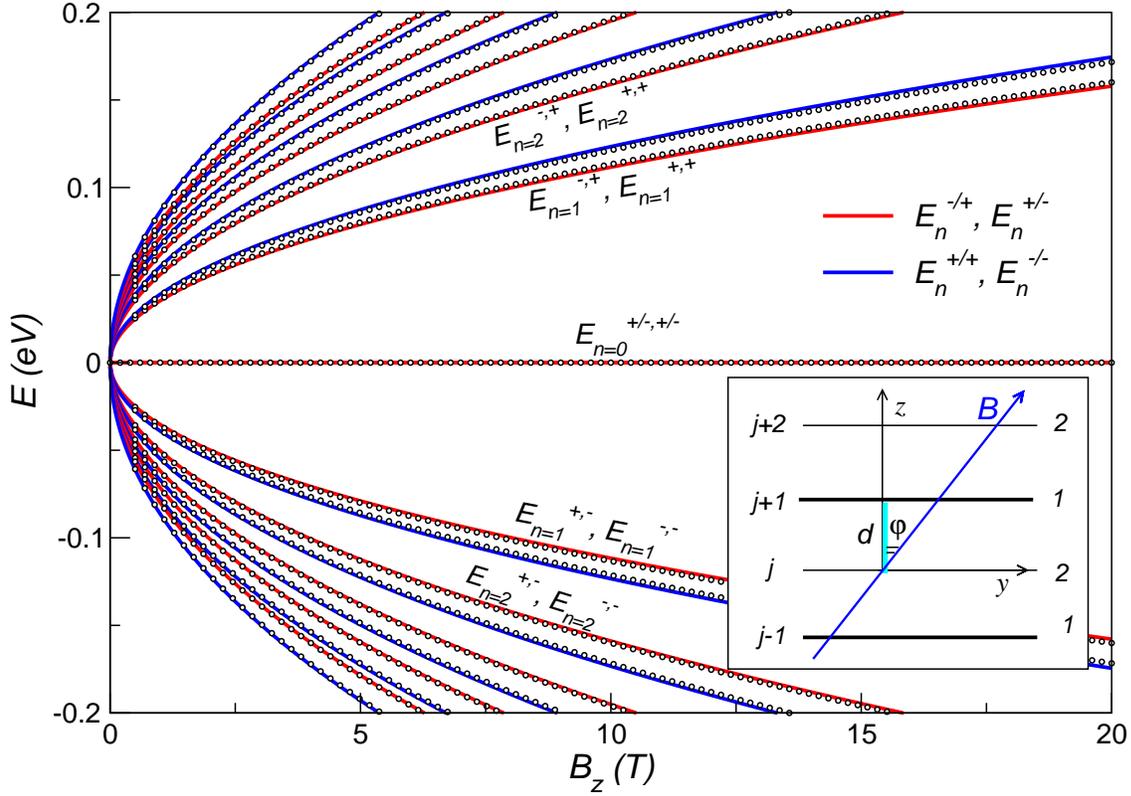}
\caption{(Color online) Splitted by tilting the magnetic field with 
$\varphi = 20^o$ LLs of graphite at the $H$ point, $E_{n}^{\pm,\pm}$, 
as a function of the perpendicular component of the magnetic field,
$B_z$.  The dotted lines correspond to results of numerical calculations.}
\label{Fig5}
\end{figure}
%\\\\\\\\\\\\\\\\\\\\\\\\\\

The small perturbation $\eta_d$ couples the states $|n\rangle_2^A$
with $|n+1\rangle_1^A$ and $|n-1\rangle_1^A$. Among them the states
$|n\rangle_2^A$ and $|n+1\rangle_1^A$ are degenerated, i.e., they
belong to the same unpertubed eigenvalues
$\pm\sqrt{\mathcal{B}(n+1)}$.  Consequently, at least the lowest order
perturbation approach suitable to remove the degeneracy must be
applied, which yields equations
\begin{eqnarray}
\left[\mathcal{B}(n+1)-E^2\right] A_{2,n}+
2 E \widetilde{\mathcal{T}}_{n,n+1} A_{1,n+1}=0,\label{eq1pp}
\\
\left[\mathcal{B}(n+1)-E^2\right] A_{1,n+1}+
2 E \widetilde{\mathcal{T}}_{n+1,n}A_{2,n}=0.\label{eq2pp}
\end{eqnarray}
The secular equation derived from Eqs.~(\ref{eq1pp}, \ref{eq2pp}) reads
\begin{equation}
\left[\mathcal{B}(n+1)-E^2\right]^2 - 4 E^2 
\widetilde{\mathcal{T}}^2_{n,n+1}=0,
\label{sec-eq}
\end{equation}
and from here we get the four eigenenergies
%\begin{eqnarray}
%E_{n+1}^{\pm,\pm}=&\pm \widetilde{\mathcal{T}}_{n,n+1}\pm
%\sqrt{\mathcal{B}(n+1)+\widetilde{\mathcal{T}}^2_{n,n+1}},\\
% &\,n = 0,1,2,\cdots.\nonumber
%\label{LLs-Hpoint}
%\end{eqnarray}
\begin{equation}
E_{n+1}^{\pm,\pm}=\pm \widetilde{\mathcal{T}}_{n,n+1}\pm
\sqrt{\mathcal{B}(n+1)+\widetilde{\mathcal{T}}^2_{n,n+1}},
\quad n = 0,1,2,\cdots
\label{LLs-Hpoint}
\end{equation}

The eigenenergies,  $E_{0}^{\mp}$ and $E_{0}^{\pm}$, steming from 
$E_{1,0}^{\pm}$, remain the same as in the perpendicular magnetic
field. In that case the degeneracy is not removed.

Lifting of LL degeneracy by the tilted magnetic field in LLs with
$n>0$ is shown in Fig.~\ref{Fig5}.  The LL splitting is of the
order of several meV, and it grows with the tilt angle, i.e., with the
in-plane magnetic field component, $B_y$.

The corresponding eigenfunctions calculated with the same level of
accuracy are presented in TABLE \ref{Table2}. They are mixed from
wave functions of both layers with an equal weight.

To test the accuracy of the above approximations we have calculated
the eigenvalues numerically with the larger basis
$|n\rangle_1^A, |n+1\rangle_1^A, |n+2\rangle_1^A, 
|n\rangle_2^A, |n+1\rangle_2^A, |n+2\rangle_2^A$
instead of the minimal one $|n+1\rangle_1^A, |n\rangle_2^A$. At the
$H$ point we have found only negligible quantitative corrections 
to the results obtained analytically, as presented in
Fig.~\ref{Fig5}. Similarly, including the higher order expansion in
$\eta_d$ does not influence the results for the chosen range of angles
and magnetic fields. 

\begin{figure}[p]
\centering
\includegraphics[width=0.95\linewidth]{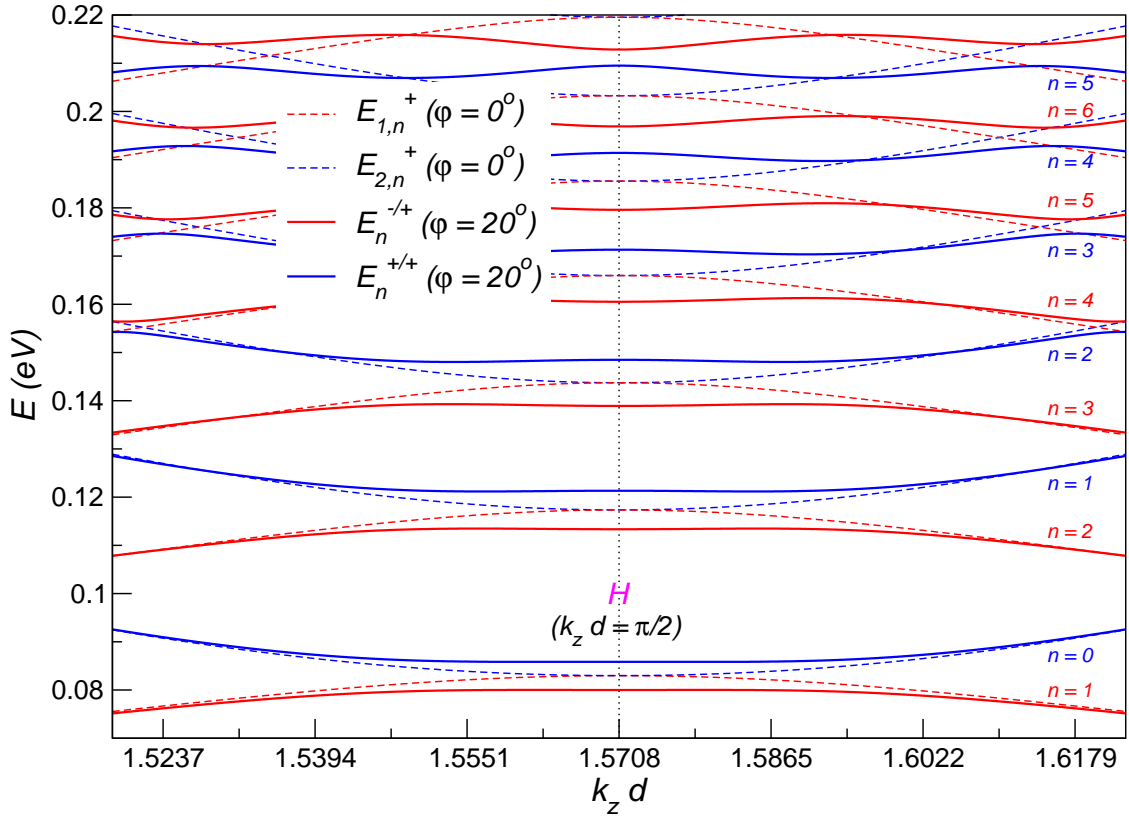}
\caption{(Color online) Graphite Landau subbands, depicted 
by the rectangular box in Fig.~\ref{Fig2}, as a function of
$k_z$ nearby the $H$ point. Dashed curves are graphite LLs, 
$E_{1,n}^{+}$ and $E_{2,n}^{+}$, in the perpendicular magnetic 
field $B_z = 5$~T, $\varphi =0^{\circ}$. 
Solid  curves are LLs of graphite, $E_{n}^{-/+}$ and 
$E_{n}^{+/+}$, subject to the tilted magnetic field with the
perpendicular component $B_z = 5$~T and the tilt angle $\varphi = 20^{\circ}$.}
\label{Fig6}
\end{figure}

The above basis allows to calculate also the $k_z$ dependence in 
the vicinity of the $H$ point defined roughly by the rectangle 
in Fig.~\ref{Fig2}. The most interesting feature is the development
of additional local extrema near the $H$ point, which are more
pronounced for LLs with higher $n$. The same is true for minigaps open
at the crossing points of LLs, as mentioned in the previous
paragraphs. The results are shown in Fig.~\ref{Fig6}.

%%%%%%%%%%%%%%%%%%
\begin{table}[p]
\caption{Energies and wave functions in tilted magnetic fields 
at the graphite $H$ point}
\begin{center}
\begin{tabular}{|c|c|}
\hline &\\ Energy & \{$\phi_1^A,\,\,\,\,\,\,\phi_1^B,\,\,\,\,\,\,\,\phi_2^A,
\,\,\,\,\,\,\,\phi_2^B$\}\\ \hline\hline
%\hline &\\ Energy & Wave function\\ \hline\hline
&\\ $E_n^\mp$ & $\{|n\rangle_1^A,
-|n-1\rangle_1^B,-|n+1\rangle_2^A,|n+2\rangle_2^B\}$\\ &\\ \hline 
&\\ $E_n^{--}$ &
$\{|n\rangle_1^A,|n-1\rangle_1^B,
-|n+1\rangle_2^A,-|n+2\rangle_2^B\}$\\ &\\ \hline
&\\ $E_n^{++}$ & $\{|n+1\rangle_1^A,-|n\rangle_1^B,
|n\rangle_2^A,-|n+1\rangle_2^B\}$\\ &\\ \hline 
&\\ $E_n^{\pm}$
& $\{|n+1\rangle_1^A,|n\rangle_1^B,
|n\rangle_2^A,|n+1\rangle_2^B\}$\\ &\\ \hline
\end{tabular}
\end{center}
\label{Table2}
\end{table}
%%%%%%%%%%%%%%%%%%%

%////////////////////////
%\\\\\\\\\\\\\\\\\\\\\\\\\\

In general, our approach must fail for magnetic fields close to the
in-plane orientation, $B_y \gg B_z$, as $\eta_d \rightarrow \infty$
for $B_z \rightarrow 0$, and the expansion in powers of $\eta_d$ is no
longer acceptable.

Also the perturbation theory is less appropriate for states with large
$n$, as the energy difference between neighboring LLs is smaller then
for states with small $n$ and, moreover, the interlayer coupling
matrix elements, $\widetilde{\mathcal{T}}_{n,n\pm 1}$, increase with
$\sqrt{n}$.  The limits of the numerical approach are not clear at
present, but we should have in mind that from the experimental point
of view the angles with almost in-plane orientation are not so
interesting due to the mosaic structure of most graphite crystals.

%*******
\section{Conclusion}
%*******
Based on the simple nearest-neighbor tight-binding quantum mechanical
model, we presented the calculation method of the band structure of
Bernal-stacked graphite subject to tilted magnetic fields. We applied
the lowest order perturbation theory to obtain analytic solutions of
the formulated equations, the accuracy of which was later checked by
the simplified numerical calculation. The special attention has been
paid to the field dependence of the LLs at the $K$ and $H$ points of
the graphite Brillouin zone where the density of states exhibits van
Hove singularities in the perpendicular magnetic field. We have found
that at the $K$ point, where the electron structure in the
perpendicular magnetic field reminds strongly that of the bilayer
graphene, the influence of the in-plane component of the magnetic
field is negligible. On the other hand, at the $H$ point, where the
electron structure mimics the behavior of the Dirac fermions, the
application of the tilted magnetic field leads to the splitting of
LLs. This splitting is of the order of several meV, which is an
experimentally observable value, and it grows with increasing of the in-plane
component of the magnetic field.

%*******
\section{Acknowledgements}
%*******
The authors benefited from discussions with Milan Orlita.  The support
of the European Science Foundation EPIGRAT project (GRA/10/E006), AV
CR research program AVOZ10100521 and the Ministry of Education of the
Czech Republic project LC510 is acknowledged.
\\
%%%%%%%%%%%%%%%%%%%%%%%%%%%%%%%%%%%%%%%%%%%%%%%%%%%%%%%%%%%%%%%%%%%%%
%%%%%%%%%%%%%%%%%%%%%%%%%%%%%%%%%%%%%%%%%%%%%%%%%%%%%%%%%%%%%%%%%%%%%

\end{document}